\documentclass[tightenlines,11pt,eqsecnum,amsmath,amssymb,aps,prd,superscriptaddress,nofootinbib]{revtex4}
\usepackage{amstext,amsfonts}
\usepackage{epsfig}
\usepackage{mathrsfs}
\usepackage{color}
\usepackage{hyperref}
\usepackage{graphics}
\usepackage{footnote}

\begin{document}
\title{Einstein Static Universe and its Stability in Generalized Rastall Gravity}
\author{Hamid Shabani}\email{h.shabani@phys.usb.ac.ir}\affiliation{Physics Department, Faculty of Sciences, University of Sistan and Baluchestan, Zahedan, Iran}\affiliation{School of Astronomy, Institute for Research in Fundamental Sciences (IPM)
P. O. Box 19395-5531, Tehran, Iran}
\author{Amir Hadi Ziaie}\email{ah.ziaie@maragheh.ac.ir}\affiliation{Research Institute for Astronomy and Astrophysics of Maragha (RIAAM), University of Maragheh, P.O. Box 55136-553, Maragheh, Iran}
\author{Hooman Moradpour}\email{hn.moradpour@maragheh.ac.ir}\affiliation{Research Institute for Astronomy and Astrophysics of Maragha (RIAAM), University of Maragheh, P.O. Box 55136-553, Maragheh, Iran}

%
\begin{abstract}
The Einstein static (ES) state is a good candidate for describing the very early universe in terms of a regular cosmological model in which the Big Bang singularity is avoided. In the present study we propose an ES solution in the framework of generalized Rastall gravity (GRG), a modified version of original Rastall theory in which the coupling parameter is allowed to vary with respect to the spacetime coordinates. Introducing an ansatz for the Rastall parameter, existence and the corresponding stability of the solutions are investigated. We show that the GRG is capable of describing a stable singularity-free state for the universe. The problem of transition from an ES to an inflationary state is also addressed. We find that a time variation of the equation of state parameter from values lying in the range $|w|<1/3$ to the value $w=-1/3$ can give rise to such a phase transition. The vector and tensor perturbations around the ES solution are studied, as well. In the case of GRG, the vector perturbations remain frozen, nevertheless, the tenor perturbations can grow in such a way that the ES solution remains stable provided the ratio of Rastall gravitational constant to the Einsteinian one always exceeds a minimum value for each tensor mode.
\end{abstract}

\maketitle
\section{Introduction}\label{int}
Einstein's general theory of relativity (GR) is a key theoretical development for 20th-century astronomy and cosmology which has led to an interesting explanation of the origin of the universe and how it has evolved~\cite{einstein1916}. In the framework of GR, the first universally accepted theory of cosmology, i.e., the hot Big-Bang cosmology~\cite{lemaitre1927} was founded and now, it is regarded as the standard cosmology which provides an opportunity to understand the cosmos exploiting elegant mathematical and physical rules~\cite{hubble1929,penzias1965}. In the context of standard model of cosmology (SMC), the universe has emerged and expanded from an extremely dense and hot primeval state. The SMC has been the starting point of studying the physics of the universe as it is later understood that it has not provided a flawless framework. However, there exist a couple of issues that SMC cannot give rational solution to comprehend them. Two famous problem are called the ``horizon problem" and the``flatness problem" which in a simple term, the former states that, why is the present universe so homogeneous and the latter questions about the observed spatially flatness of the universe~\cite{weinberg2008}. Another, key problem which is observationally confirmed seeks a physical mechanism to explain the observed accelerated expansion of the universe. Some of these observations are type Ia supernovae (SNIa)~\cite{riess1999,tonry2003,nnop2003}, large-scale structure~\cite{tegmark2006}, baryon acoustic oscillations~\cite{perciva2007}, the cosmic microwave background radiation (CMBR)~\cite{Komastu2009,akrami2020}, and weak lensing~\cite{jain2003}.
\\\\
Soon, it was realized that problems with initial conditions existed in the SCM can mostly be solved by intervening a rapid accelerated expansion phase which must be occurred at the beginning of the hot Big-Bang era. The technical expression for the mentioned mechanism is the so-called cosmic ``inflation'' which has historically been proposed in different contexts e.g., within singularity problems~\cite{starobinsky1980} and as a cosmological phase transition~\cite{sato1981}. Nevertheless, the necessity of an inflationary era to solve problems with initial conditions is highlighted in~\cite{guth1981,linde1982,albrecht1982}. In addition to dealing with the initial conditions, inflation can also explain how structures have been formed from ancient cosmological seeds~\cite{mukhanov1981}. Notwithstanding that the idea of inflation has been successful to explain some of main issues of the SCM, it suffers from another problem which is related to initial singularity of the universe i.e., the universe necessarily had a singular beginning for generic inflationary cosmologies~\cite{BordeVil}. Up to now, different resolutions of the problem have been proposed that the most of which originate from quantum gravity. Some of these proposals are for instance, the idea of pre-Big-Bang~\cite{lidsey2000,gasperini2003} and cyclic scenarios~\cite{steinhardt2002,khoury2004}, bouncing cosmology~\cite{ijjas2018} and emergent Universe (EU) scenario~\cite{ellis2004}.
\\\\
The idea of EU scenario interestingly tries to present an inflationary period without an initial singularity. In an EU model there is no timelike singularity prior to the inflationary era, instead, the universe eternally exists in an stable static state which is called Einstein static (ES) state and then evolves to an accelerated speed-up stage. Therefore, EU scenario states that the universe would evolve from an ES phase (instead of a Big-Bang singularity) to an inflationary state. Within the past decades many authors have investigated ES solutions in the context of different gravity theories. First studies in the framework of GR have been performed by Eddington in 1930~\cite{eddington1930}. He found that ES solutions bear some instability against homogeneous and isotropic perturbations.
Within the same framework, a model of closed universe including ultra-relativistic matter and a cosmological constant, which asymptotically meets the ES model, has been introduced by Harrison~\cite{harrison1967}. However, his model could not exit to an inflationary phase.
In~\cite{gibbons1987,gibbons1988,barrow2003} it has been shown that a universe filled with a perfect fluid has a neutral stability with respect to all types of linear perturbations if $c_{s}^{2}>1/5$, where $c_{s}$ being the sound speed. Also, existence of ES solutions has been investigated in the context of theories beyond GR that some of interesting works are as follows: the ES universe containing a minimally coupled scalar field $\phi$ with a self-interacting potential $V(\phi)$ has been considered in ~\cite{ellis2004}. Effects of an exotic matter on the ES solution have been studied in~\cite{mukherjee2006}. The ES universe has been also investigated in brane-world scenarios~\cite{lidsey2006,banerjee2008,zhang2014}, Einstein-Gauss-Bonnet
theory~\cite{mukerji2010,paul2010}, $f (R)$ gravity~\cite{barrow1983,goswami2008,seahra2009},  $f (T)$ theory~\cite{wu2011}, loop quantum cosmology~\cite{canonico2010,bag2014}, $f(R,T)$ gravity~\cite{shab2017} and in Einstein-Cartan-Brans-Dicke gravity~\cite{shab2019}.
\par
The conservation of energy momentum tensor (EMT) has usually been assumed in the most investigations based on some physical facts. However, it was turned out that the conservation of EMT may be violated in quantum systems~\cite{birrell1982,koivisto2006,minazzoli2013}. Besides, Rastall in 1972 has emphasized that there is no observational and theoretical evidence to prohibit the non-conservation of EMT in gravitational physics~\cite{rastall1972}. He justified his proposal by this fact that the conservation of EMT (which is mathematically stated as $T^{\mu\nu}_{\ \ \ ;\mu}=0$) has only been examined in a flat spacetime or in the presence of weak gravitational fields. Motivated by this idea, Rastall then assumed that the covariant divergence of the EMT is no longer zero but, proportional to the covariant derivative of the Ricci curvature scalar, i.e., $T^{\mu\nu}_{\ \ \ ;\mu}=\lambda R^{;\nu}$. The constant $\lambda$ is considered as a measure of non-minimal coupling between matter and geometry. We note that in the weak field limit, the usual conservation law is recovered. Recently, a generalization of Rastall gravity has been proposed in which the Rastall constant parameter is replaced by a function of arbitrary spacetime coordinates thus the modified EMT conservation reads $T^{\mu\nu}_{\ \ \ ;\mu}=\left(\lambda R\right)^{;\nu}$. Interestingly, such a modification can explain the present accelerated expansion of the universe~\cite{moradpour2017}. Also it has been shown that a coupling between the energy-momentum source and geometry can produce the primary inflationary expansion even in the absence of a matter source. We refer to this extension of Rastall gravity as the generalized Rastall gravity (GRG). 

Different features of GRG have been studied, up to now, among which we can quote: the authors of~\cite{das2018} obtained pre-inflationary and late time accelerated expansion solutions. In~\cite{lin20192} another extension of the Rastall idea has been proposed. In the mentioned work an arbitrary function of $R$, i.e., $f(R)$ has been utilized instead of the Ricci scalar. Two of the present authors tried to introduce a Lagrangian approach for GRG by connecting GRG to $f(R,T)$  gravity~\cite{shabani2020}. In~\cite{moradpour2021}, the authors have shown that GRG is compatible with a running gravitational coupling and thus with Dirac hypothesis, in this case, the present dark energy behavior of the universe can be understood. Also, GRG has been discussed within the dynamical system approach in~\cite{shabani2022}. Using this method, it has been illustrated that a proper radiation, dark matter and dark energy sequence can be achieved in the GRG framework.
\par
In the present work we discuss the solutions of ES universe in the context of GRG. In Sect.~\ref{sec2} the field equations of GRG and necessary equations will be presented. In Sect.~\ref{sec3} we show that not only in GRG framework stable ES solutions can be found but also a successful exit from an everlasting ES state to an inflationary era is possible. In Sect.~\ref{sec4} the stability of our solutions are evaluated against the scalar, vector and tensor perturbations. Our conclusions are drawn in Sect.~\ref{conc}.

\section{A concise description of field equations of GRG}\label{sec2}
In the present section we briefly describe the main field equations of GRG along with necessary relations. As mentioned within the introduction, the GRG model begins with the following condition on the EMT
\begin{eqnarray}\label{grg0}
T^{\mu\nu}_{\ \ \ ;\mu}=\left(\lambda R\right)^{;\nu},
\end{eqnarray}
where $R$ and $T_{\mu\nu}$ denote the Ricci scalar and the EMT, respectively. Also, $\lambda$  is called the Rastall parameter which can be an arbitrary function of the spacetime coordinates. It is worth mentioning that, recently, a novel generalization of GRG has been introduced in which the condition on EMT conservation is considered as $T^{\mu\nu}_{\ \ \ ;\mu}=\mathscr{A}^{\mu\nu}_{\ \ \ ;\mu}$ with $\mathscr{A}_{\mu\nu}=\lambda g_{\mu\nu}f(R,T^\mu_{\,\,\,\mu})$, in order to study the evolution of dark energy in GRG~\cite{lin2020}.
\par
From Eq.~(\ref{grg0}), the following field equation can be easily obtained 
\begin{eqnarray}\label{grg1}
G_{\mu\nu}+\kappa\lambda g_{\mu\nu}R=\kappa T_{\mu \nu},
\end{eqnarray}
where $G_{\mu\nu}$ is the Einstein tensor and $\kappa$ is the Rastall gravitational coupling constant. Hence, Eqs.~(\ref{grg0}) and (\ref{grg1}) are the principal field equations of GRG and based on the type of application, different forms of Rastall parameter have been introduced up to know. In~\cite{das2018} this parameter has been chosen as $\lambda=(1+d_{0}H)/3\kappa(w+1)$\footnote{For the sake of convenience, the model parameters have been rewritten according to notations of the present paper.} with $d_{0}$ being a constant, in order to discuss some cosmological implications of GRG. The authors of~\cite{moradpour2021} have shown that the Dirac proposal can be achieved considering the coupling parameter as $\lambda=\zeta H^{n}/R$, where $n$ is an arbitrary constant. Also in~\cite{shabani2022}, it has been shown that the same Rastall parameter can be responsible for current accelerated expansion of the universe. In present study we use a homogeneous power-law form for the Rastall parameter as
\begin{eqnarray}\label{grg2}
\lambda(t)=\frac{\zeta}{\kappa} a(t)^{n},
\end{eqnarray}
where $a(t)$ is the scale factor and $\zeta$ is a constant. We use the EMT of a perfect fluid
\begin{align}\label{grg3}
T_{\mu\nu}=(\rho+p)u_\mu u_\nu + p g_{\mu\nu},
\end{align}
as the matter content of the universe, where $\rho$, $p$ and $u_\alpha$ being the matter density, pressure and four-vector velocity of the fluid, respectively. The line element for ES model is taken as a non-flat FLRW metric
\begin{align}\label{grg4}
ds^{2}=-dt^{2}+a^{2}(t) \left [\frac{dr^{2}}{1-kr^2}+r^{2}d\Omega^2\right ],
\end{align}
where $k$ and $d\Omega^2$ denote the spatial curvature and line element on a unit two-sphere, respectively. The equations of motion along with conservation equation then read
\begin{align}
&3(1-4\kappa\lambda)H^2 - 6\kappa\lambda\dot{H}+3(1-2\kappa\lambda)\frac{k}{a^{2}}=\kappa\rho,\label{grg5}\\
&3(1-4\kappa\lambda)H^2 +2(1-3\kappa\lambda)\dot{H}+(1-6\kappa\lambda)\frac{k}{a^{2}}=-\kappa p,\label{grg6}\\
&\frac{d}{dt}\left(\rho+\lambda R\right)+3H(\rho+p)=0.\label{grgcons}
\end{align}
We note that the above set of equations are not independent, namely, Eq.~(\ref{grgcons}) can be obtained by differentiating Eqs.~(\ref{grg5}) and (\ref{grg6}). We therefore consider only these two equations. Hence, for a barotropic perfect fluid with $p=w\rho$, eliminating $\rho$ from Eqs. (\ref{grg5}) and (\ref{grg6}) gives
\begin{align}\label{grg7}
3(1+w)(1-4\kappa\lambda)H^2+2\big[1-3(1+w)\kappa\lambda\big]\dot{H}+3\Big[1+3w-6(1+w)\kappa\lambda\Big]\frac{k}{a^{2}}=0.
\end{align}
\\
In the next section we obtain the Einstein static (ES) solution and investigate the corresponding stability properties. Also, different possibilities to exit from the ES state will be discussed.
\section{ES solutions from dynamical system approach and a successful exit}\label{sec3}
In the present section we seek for possible ES solutions using Eq.~(\ref{grg7}). To this aim, we define two variables $x(t)=a(t)$ and $y(t)=\dot{a}(t)$ in order to construct a closed dynamical system. With the help of these definitions one finds $H=y/x$ and $\dot{H}=\dot{y}/x-(y/x)^2$. Substituting for $H$ and $\dot{H}$ into Eq.~(\ref{grg7}) the following autonomous system can be obtained
\begin{align}
&\dot{x}=y,\label{grg8}\\
&\dot{y}=-\frac{ f(\kappa\lambda;w)\big[y^2+3kx^{2}\big]}{g(\kappa\lambda;w)x},\label{grg9}
\end{align}
where we have defined
\begin{align}
&f\left(\kappa\lambda;w\right)=1+3w-6(1+w)\kappa\lambda,\nonumber\\
&g\left(\kappa\lambda;w\right)=2\left[1-3(1+w)\kappa\lambda\right].\nonumber
\end{align}
\\
We firstly study possible stable ES solutions in the subsequent subsection and in subsection (\ref{sec3.2}) we discuss the conditions for which the universe can undergo a smooth exit from a stable ES solution to an eternal inflationary phase.
\subsection{The ES solutions}\label{sec3.1}
\begin{figure}[h!]
\begin{center}
\epsfig{figure=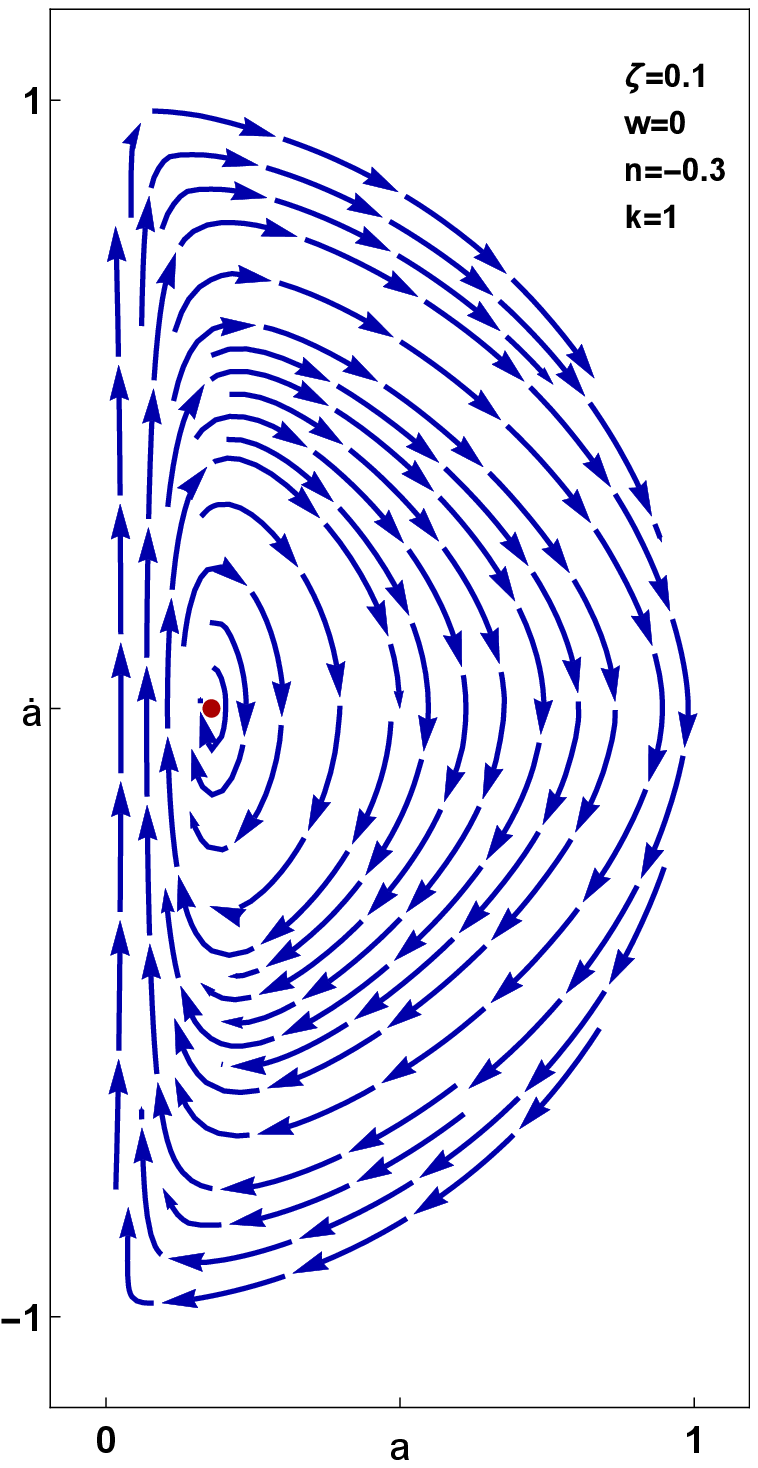,width=5.cm}\hspace{2mm}
\epsfig{figure=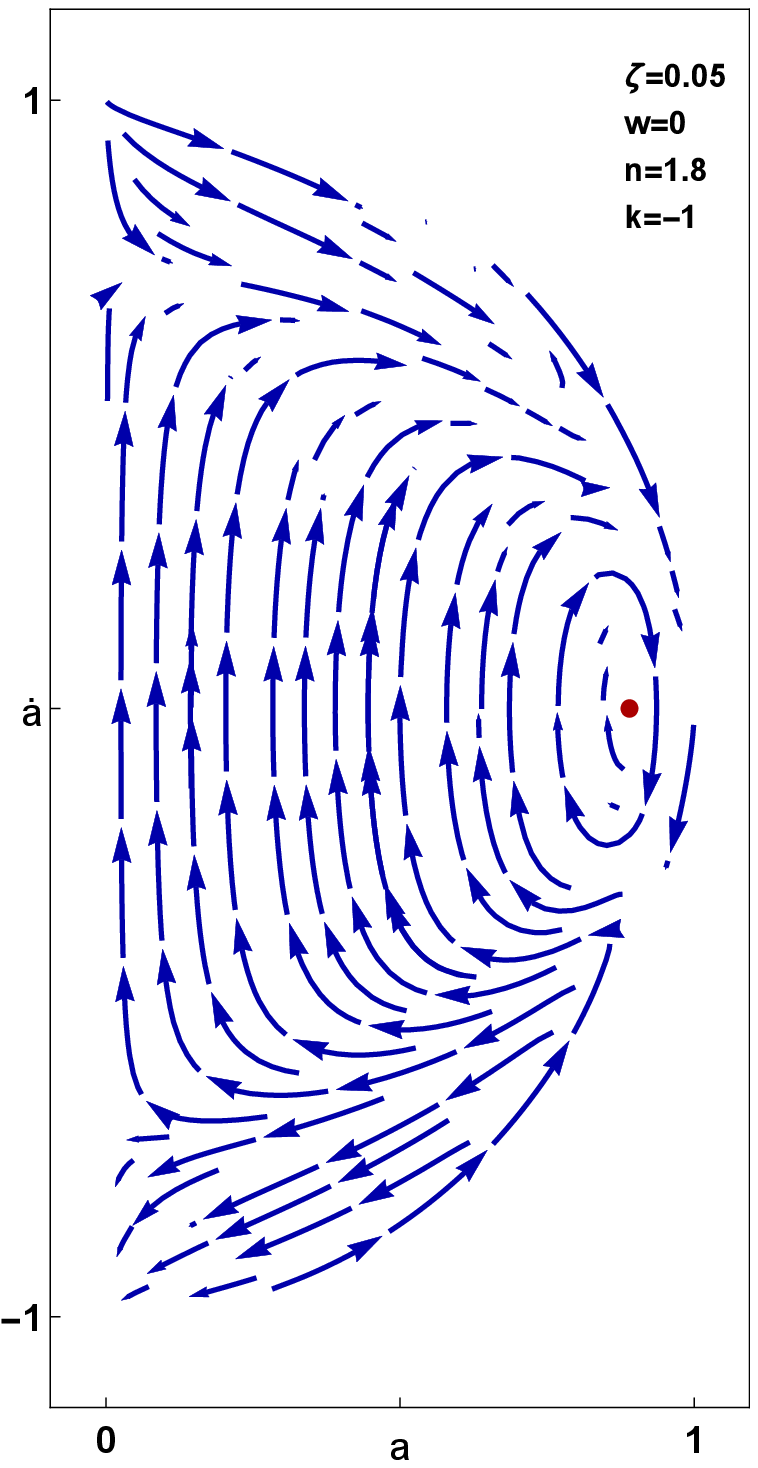,width=5.cm}
\caption{Phase space portraits of GRG related to system of eqs. (\ref{grg8})-(\ref{grg9}) and ansatz (\ref{grg2}) for two different sets of model parameters.}\label{fig1}
\end{center}
\end{figure}

\begin{figure}[h!]
\begin{center}
\epsfig{figure=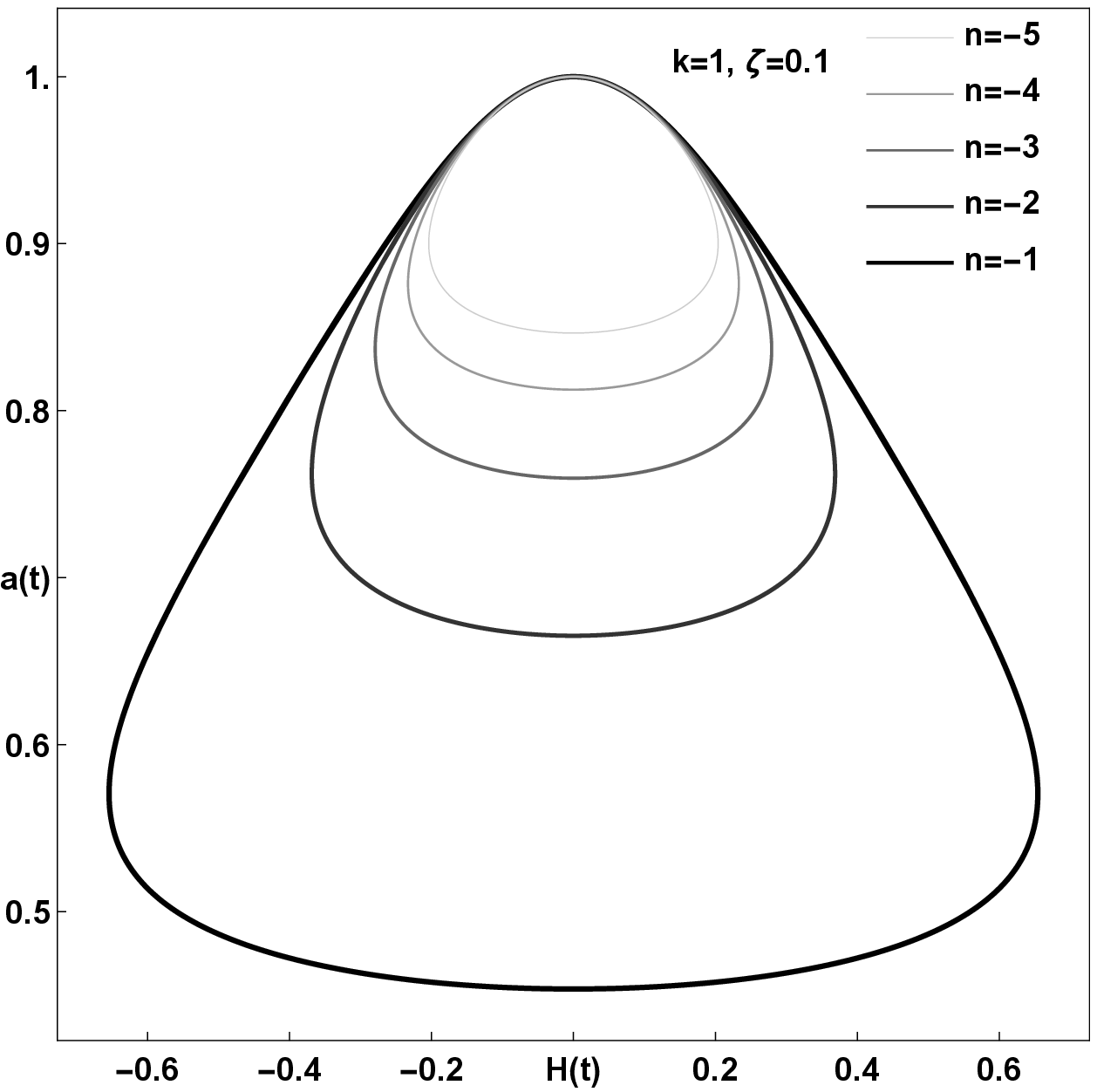,width=8.cm}\hspace{2mm}
\epsfig{figure=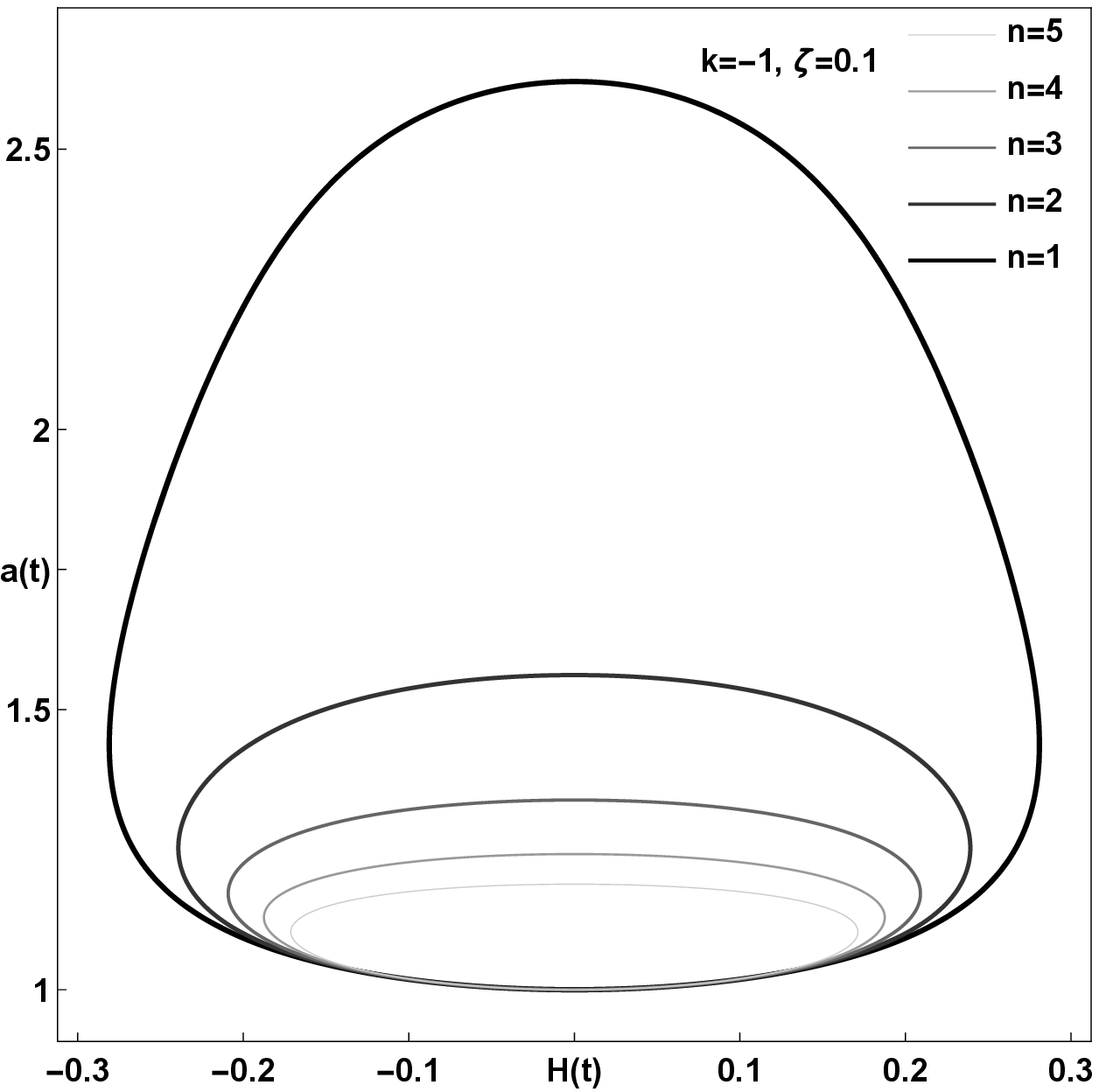,width=8.cm}
\caption{Numerical solutions of Eqs.~(\ref{grg8})-(\ref{grg9}) in $(H,a)$ plane for both closed (left panel) and open (open panel) FLRW universes. Initial values $a_{\rm ES}=1$ and $\dot{a}_{\rm ES}=0$ has been chosen for both panels.}\label{fig2}
\end{center}
\end{figure}

The ES solution can be obtained by setting the time derivatives of different variables to zero. The system (\ref{grg8})-(\ref{grg9}) for coupling parameter (\ref{grg2}), has only one fixed point solution given by
\begin{align}\label{grg10}
&x_{\rm ES}=\left[\frac{1+3w}{6(1+w)\zeta}\right]^{\frac{1}{n}}=\mathcal{S}^{\frac{1}{n}},~~~~~~\zeta~\&~n\neq0,\\
&y_{\rm ES}=0.\nonumber
\end{align}
As can be seen, the radius of the ES universe (note that $x_{\rm ES}=a_{\rm ES}$) depends only on the Rastall parameter (via constants $n$ and $\zeta$) and the matter content of the universe (via the equation of state parameter, $w$). Note that we cannot evaluate $x_{\rm ES}$ for $\zeta=0$ or $n=0$ since these values lead to GR or $\Lambda$CDM (see eqs. (\ref{grg5})-(\ref{grg6}) and ansatz (\ref{grg2})) model, i.e. for $\zeta=0$ and $n=0$ solution (\ref{grg10}) is not valid. Next, we proceed with determining the stability properties. This aim can be achieved by evaluating the eigenvalues of Jacobian matrix $(J_{ij}=\partial\dot{q}_i/\partial q_j)$, which in our case are given by
\begin{align}\label{grg11}
\lambda_{1,2}=\pm \frac{i}{a_{\rm ES}}\sqrt{3kn\frac{3w+1}{3w-1}}.
\end{align}
We therefore observe that the fixed point solution ($x = a_{\rm ES},y =0$) with the condition that the radius of ES universe must be real and positive, is a center equilibrium point provided that
\begin{align}\label{grg12}
k=1,~\left\{
\begin{array}{l}
\zeta<0,~n>0,~~-1<w<-\frac{1}{3}\\
\zeta>0,~n>0,~~w<-1\bigvee w>\frac{1}{3}\\
\zeta>0,~n<0,~~|w|<\frac{1}{3}
\end{array}
\right.
~~~{\mbox and}~~~
k=-1,~
\left\{
\begin{array}{l}
\zeta<0,~n<0,~~-1<w<-\frac{1}{3}\\
\zeta>0,~n>0,~~|w|<\frac{1}{3}\\
\zeta>0,~n<0,~~w<-1\bigvee w>\frac{1}{3}
\end{array}
\right.
\end{align}
$\forall n, |n|\neq 1/2, 1/4, 1/6,...,$ when $\mathcal{S}>0$ holds and 
\begin{align}\label{grg13}
k=1,~\left\{
\begin{array}{l}
n<0,~~|w|<\frac{1}{3}\\
n>0,~~|w|>\frac{1}{3}
\end{array}
\right.
~~~{\mbox and}~~~
k=-1,~
\left\{
\begin{array}{l}
n<0,~~|w|>\frac{1}{3}\\
n>0,~~|w|<\frac{1}{3}
\end{array}
\right.
\end{align}
$\forall n, |n|= 1/2, 1/4, 1/6,\ldots$, when $\mathcal{S}<0$ holds. Note that in the latter case the radius of ES universe is real and positive for all values of $\zeta$ and $w$ parameters.\\\\
Phase space trajectories for which conditions (\ref{grg12})-(\ref{grg13}) are satisfied circulate around the fixed point, because the eigenvalues indicate that we have a center equilibrium critical point. In Fig.~\ref{fig1} two examples of such trajectories have been demonstrated. We see that for different set of constants (which satisfy conditions (\ref{grg12})), center equilibrium solutions have been obtained which illustrate the stability type of fixed point solution (\ref{grg10}) very well.
\\
We also depict evolutionary diagrams for different variables by numerically solving Eqs. (\ref{grg8})-(\ref{grg9}). These solutions indicate that all quantities have an oscillatory behavior around their equilibrium centers. Figure~\ref{fig2} shows various plots in $(H,a)$ plane for both a closed (left panel) and an open (open panel) universe. The equilibrium point is obtained from (\ref{grg10}) for each curve. Figure~\ref{fig2} also shows that for a closed universe the spatial extension of universe can oscillate between the radius of unity to smaller ones. On the contrary, in an open universe the radius of the universe can grow to a greater value than unity then contracts to smaller radii.

\begin{figure}[h]
\begin{center}
\epsfig{figure=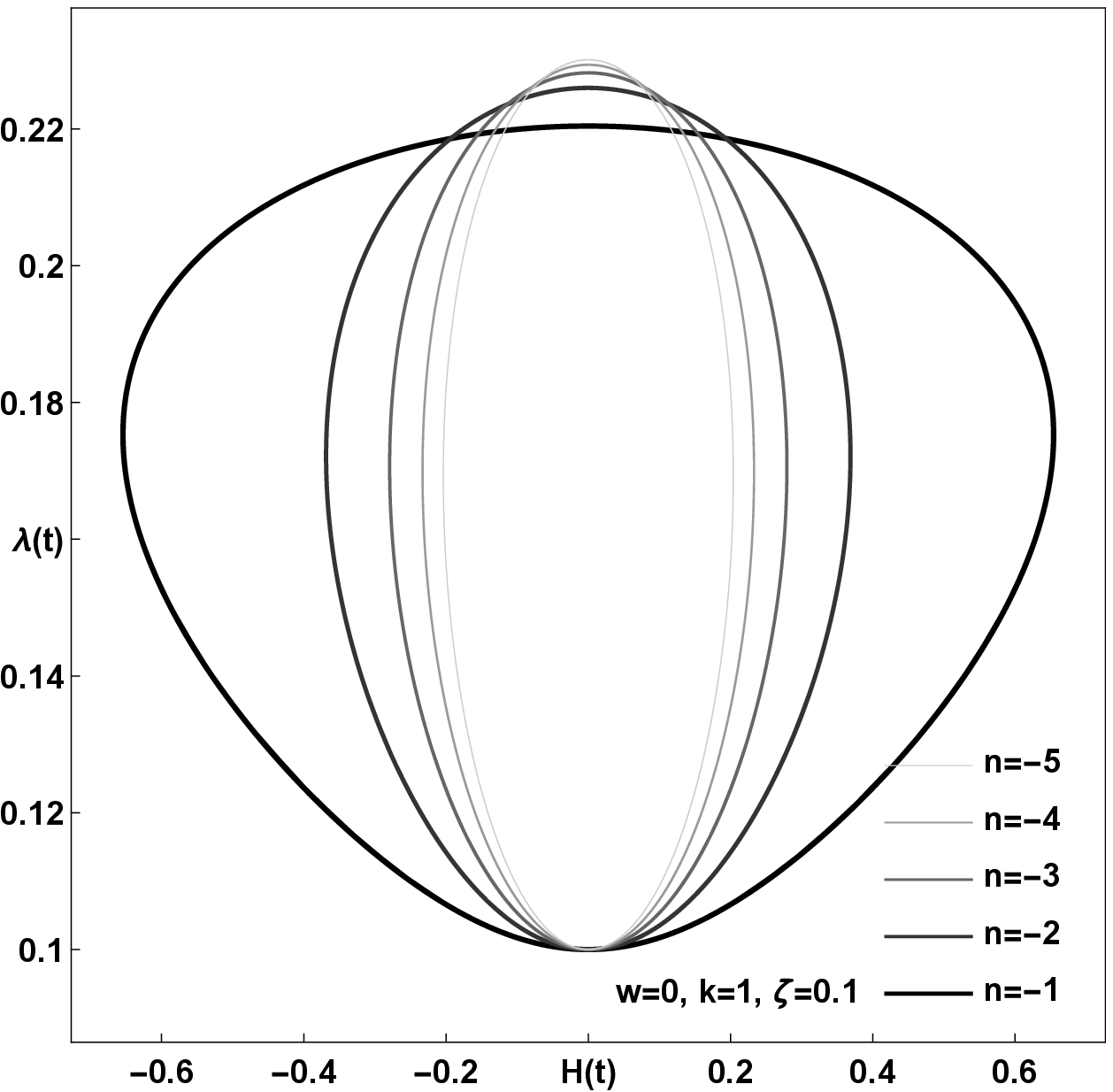,width=7.9cm}\hspace{2mm}
\epsfig{figure=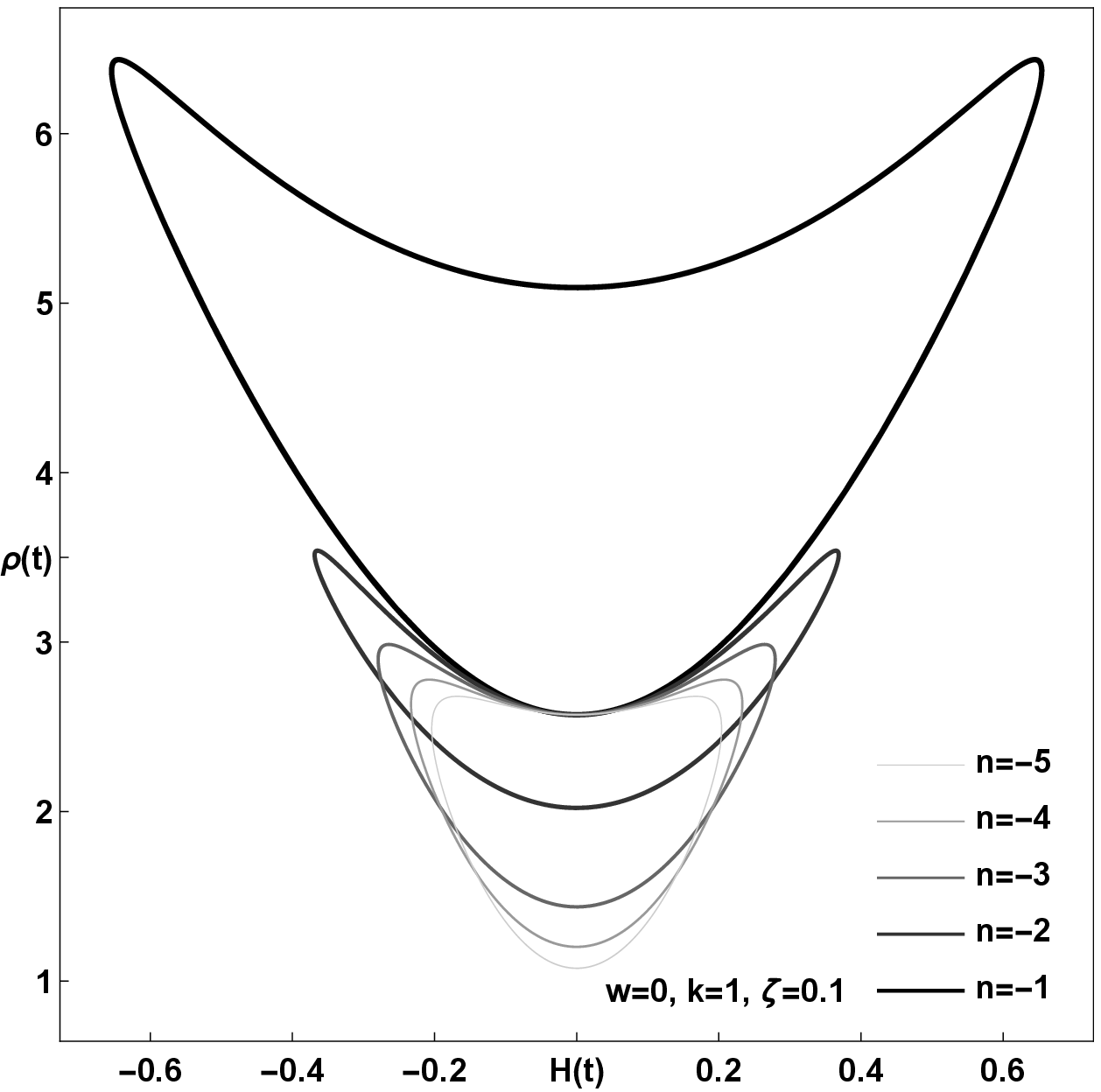,width=7.9cm}
\caption{Evolution of the Rastall and matter density parameters with respect to the Hubble parameter $H$ for the same values of constants. Initial values $a_{\rm ES}=1$ and $\dot{a}_{\rm ES}=0$ has been used for both panels.}\label{fig3}
\end{center}
\end{figure}

Figure~\ref{fig3} demonstrates the variation of Rastall parameter $\lambda$ and the matter density $\rho$ with respect to the Hubble parameter $H$ for the same values of model parameters. This figure shows that the universe behaves wavier as the absolute value of $n$ parameter increases.
\subsection{From ES to an EU state }\label{sec3.2}
\begin{figure}[h]
\begin{center}
\epsfig{figure=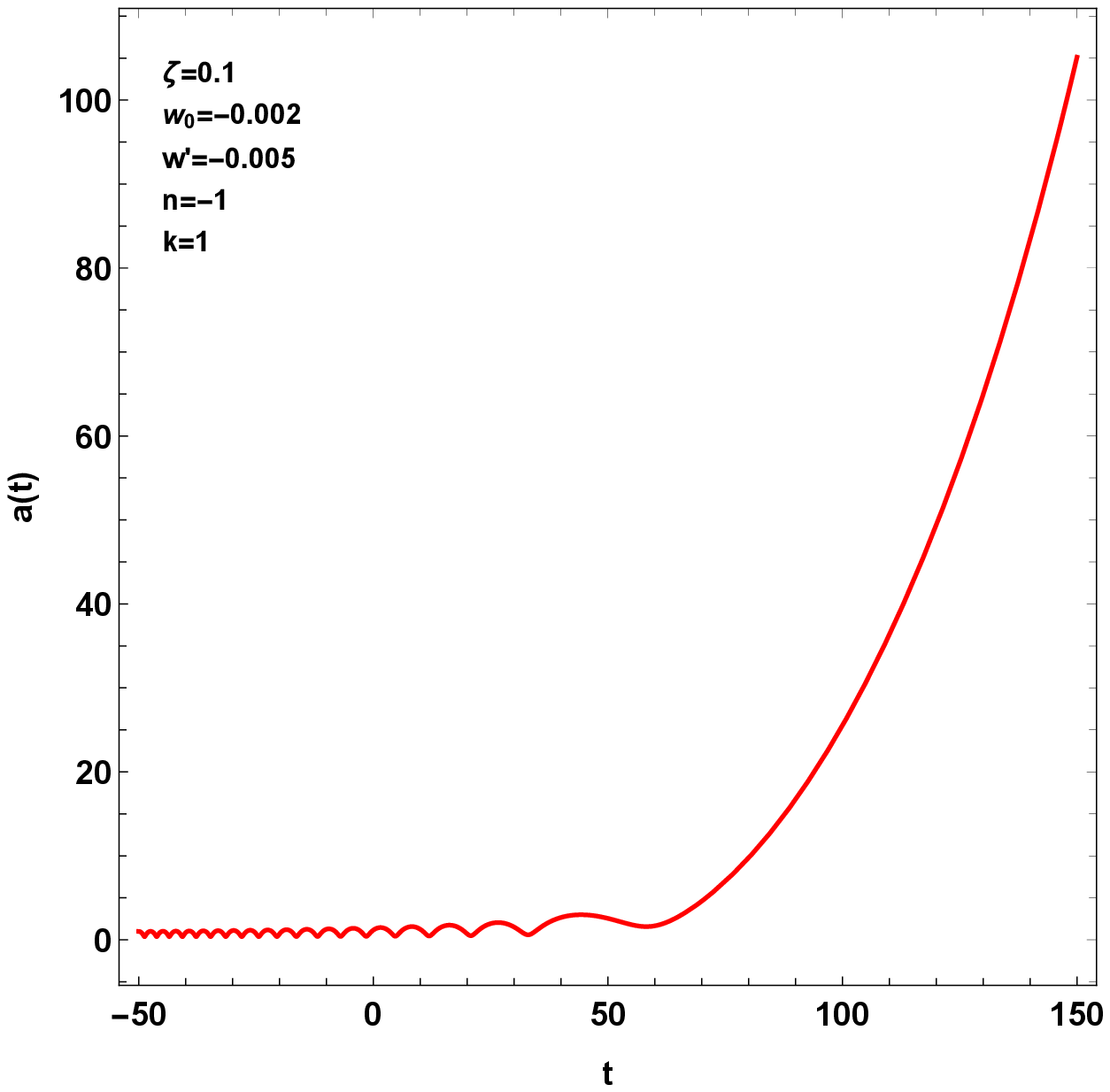,width=8.cm}\hspace{2mm}
\epsfig{figure=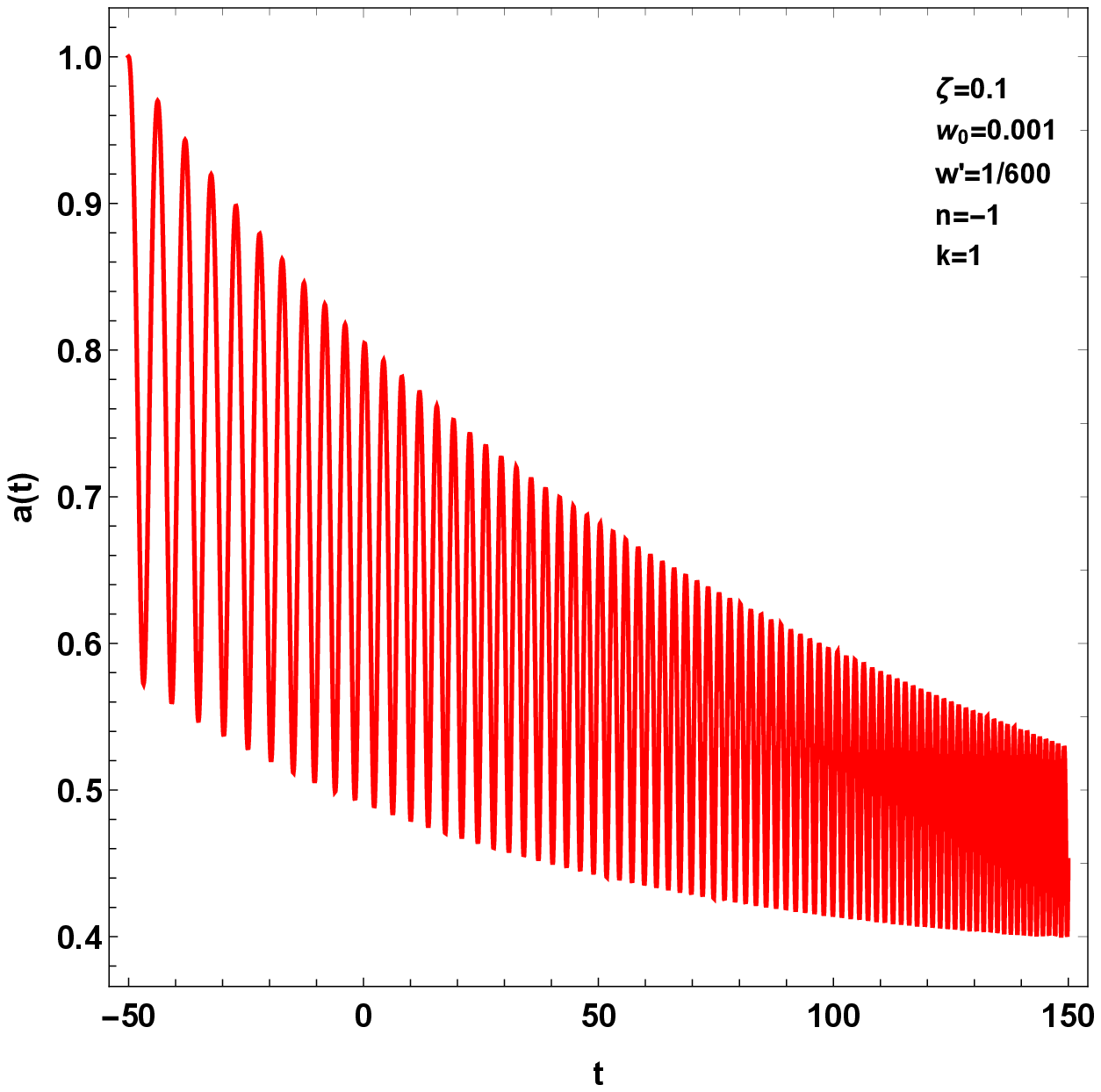,width=8.cm}\vspace{2mm}
\epsfig{figure=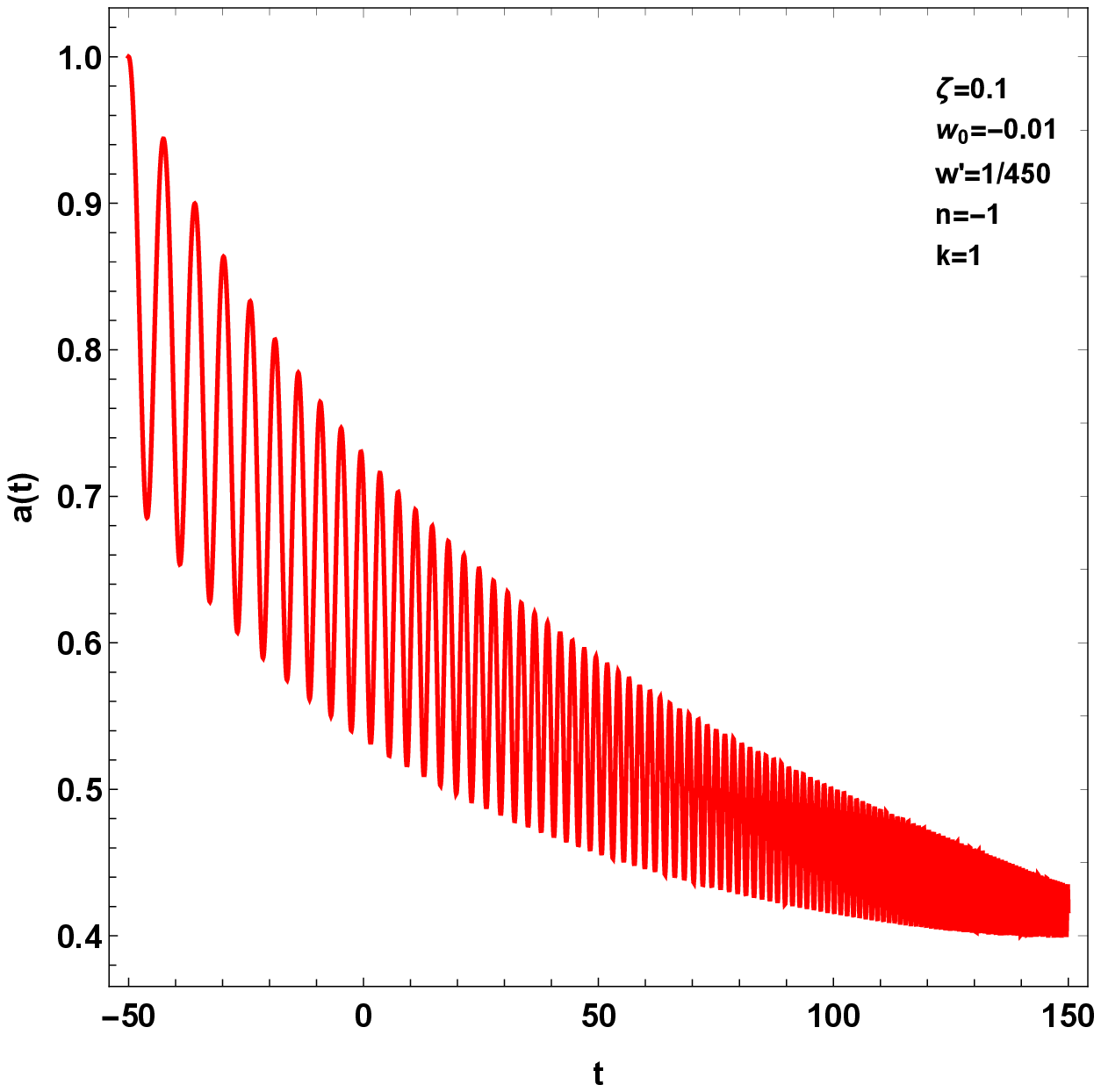,width=8.cm}\hspace{2mm}
\epsfig{figure=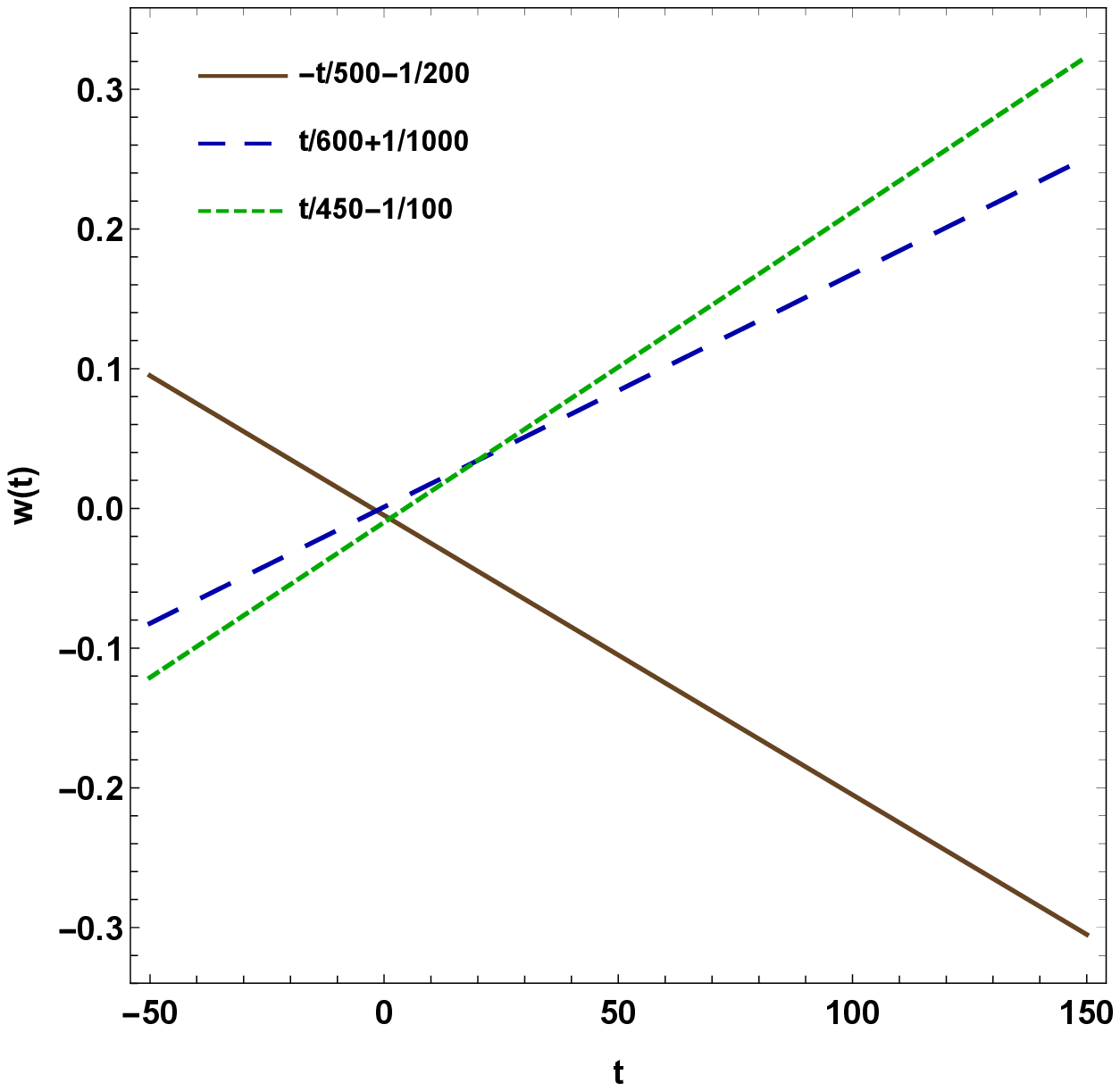,width=8.cm}
\caption{Exit from an Einstein static universe. Three different parameterizations have been considered for $w$ in lower right panel with the constants represented in other panels. We see that only a variation of $w$ to $-1/3$ can make the universe exit from the ES state.}\label{fig4}
\end{center}
\end{figure}

An interesting question in the context of ES scenarios is the possibility of exiting the universe from an ES state to an ever lasting accelerated expansion state, i.e., an EU scenario. In the present section we seek for an answer in relation to the GRG theories. The valid ranges of (\ref{grg12}) tell us that when $\zeta>0$ and $n<0$ only for $-1/3<w<1/3$ the ES universe remains in a stable balance and otherwise it is unstable. In our model, we assumed a barotropic perfect fluid which has a constant equation of state parameter. Now, we consider the situation where $w$ could vary with respect to time and check out the result. This allows us to understand what would happen if for example the dark energy production begins or the pressure of cosmic matter grows. To examine each possibility, we parameterize the equation of state parameter as $w=w't+w_{0}$ and choose three different set of constants $(w',w_{0})$ in such a way that transitions to either boundaries of the interval $-1/3<w<1/3$ could take place\footnote{Here, to consider the question of transition from an ES to an EU universe, we only consider those transitions which may be happened for $\zeta>0$ and $n<0$ and the other cases of intervals (\ref{grg12}) can also obviously be considered.}. In the lower right panel of Fig.~\ref{fig4} which is provided for a spatially closed FLRW universe, three different parametrizations for $w$ have been sketched; solid brown line for $w=-t/500-1/200$ which shows that the equation of state parameter decreases from the larger amount $-0.005$ to approximately $-1/3$. As the upper left panel shows, within this range of variation, the universe exits from the ES state to an expanding one which could be an inflationary phase. Note that, the variation of $w$ cannot pass the exact value $-1/3$ since from (\ref{grg10}) the value of $a_{\rm ES}$ becomes zero. The two other plots for the scale factor are provided for two situations in which a variation $|w|\approx0\to w\approx 1/3$ occurs. From the upper right and lower left panels we see that the universe tends to shrink to a smaller scale ES state. Therefore, a change of the cosmic matter structure from the valid range $|w|<1/3$ to $-1/3$ makes the universe to experience a phase transition from an eternal ES state to an everlasting expansion state.

\section{Scalar, vector and tensor perturbations}\label{sec4}
Now, we wish to examine whether the Einstein static universe solutions found above are stable under small perturbations. Hence, our aim in the present section is to consider the stability of linear homogeneous scalar perturbations around the ES solution and also discuss the vector and tensor perturbations. We first obtain perturbed equations for a general Rastall parameter to better observe the governing equations and then, we proceed with ansatz (\ref{grg2}). 
\subsection{Scalar perturbations}
In order to derive the equations governing homogeneous scalar perturbations, we introduce perturbed quantities for the scale factor, the matter energy density and the Rastall parameter as follows
\begin{align}\label{grg16}
a(t)=a_{\rm ES}[1+\delta a(t)],~~~~~~\rho(t)=\rho_{\rm ES}[1+\delta \rho(t)],~~~~~~\kappa\lambda(t)=\kappa\lambda_{\rm ES}[1+\delta( \kappa\lambda(t))],
\end{align}
where unperturbed quantities $a_{\rm ES}$  and $\rho_{\rm ES}$ being estimated about the equilibrium state. Note that we hereafter drop the ${\rm ES}$ subscript from all unperturbed quantities for simplicity. Also, in deriving the perturbed equations we use the unperturbed ones (\ref{grg5})-(\ref{grg6}) when all time derivative of variables vanish. Particularly, from Eqs.~(\ref{grg5})-(\ref{grg6}) for a barotropic perfect fluid we obtain 
\begin{align}\label{grg17}
\kappa \lambda_{\rm ES}=\frac{3w+1}{6(w+1)},
\end{align}
which can also be obtained from solution (\ref{grg10}). After lengthy but straightforward calculations we find non-linear scalar perturbations of the field equations, as
\begin{align}\label{grg18}
&\mathcal {A}{\ddot{\delta}a}+\mathcal{B}{\dot{\delta} a}^2-6\Big[ \mathcal{C}\delta a+\mathcal{D}\delta\lambda+\mathcal{E}\delta a \delta\lambda\Big]\frac{k}{a^2}=\kappa\rho\delta\rho,\\
&\mathcal {A}=-6\kappa\lambda(1+\kappa\delta\lambda)(1-\delta a),~~~~~\mathcal{B}=3\big[1-2\kappa\lambda(1+\kappa\delta\lambda)\big](1-2\delta a),\nonumber\\
&\mathcal{C}=1-2\kappa\lambda,~~~~~~~~\mathcal{D}=\kappa^{2}\lambda,~~~~~\mathcal{E}=-2\kappa^{2}\lambda,\nonumber
\end{align}
 and
\begin{align}\label{grg19}
&\mathcal {F}\ddot{\delta}a+\mathcal{G}{\dot{\delta} a}^2-2\Big[ \mathcal{H}\delta a+\mathcal{I}\delta\lambda+\mathcal{J}\delta a \delta\lambda\Big]\frac{k}{a^2}=-\kappa w \rho\delta\rho,\\
&\mathcal {F}=2\big[1-3\kappa\lambda(1+\kappa\delta\lambda)\big](1-\delta a),~~~~~\mathcal{G}=\big[1-6\kappa\lambda(1+\kappa\delta\lambda)\big](1-2\delta a),\nonumber\\
&\mathcal{H}=1-6\kappa\lambda,~~~~~~~~\mathcal{I}=3\kappa^{2}\lambda,~~~~~\mathcal{J}=-6\kappa^{2}\lambda,\nonumber
\end{align}
where $\dot{\delta}a=d/dt(\delta a)$ and use has been made of only the approximation $(1+x)^n\approx1+nx$ for $|x|<<1$, in order to avoid fractional terms. To obtain linearized perturbed field equations, second order perturbations like {${\dot{\delta} a}^2$}, $\delta\lambda\delta a,\ldots$, should be omitted. We then get
\begin{align}
&\kappa\lambda\ddot{\delta}a+\left[\left(1-2\kappa\lambda\right)\delta a+\kappa^{2}\lambda\delta\lambda\right]\frac{k}{a^2}=-\frac{\kappa}{6}\rho\delta\rho,\label{grg20}\\
&\left(1-3\kappa\lambda\right)\ddot{\delta}a-\left[\left(1-6\kappa\lambda\right)\delta a+3\kappa^{2}\lambda\delta\lambda\right]\frac{k}{a^2}=-w\frac{\kappa}{2}\rho\delta\rho.\label{grg21}
\end{align}
Equations (\ref{grg20})-(\ref{grg21}) are the most general ones which govern the linearized scalar perturbations in GRG for an arbitrary Rastall perturbation $\delta\lambda$. Note that in general, we have $\delta\lambda=\delta\lambda(\delta a,{ \dot{\delta}a}^{2}, \delta a\delta\lambda ...)$. Eliminating $\delta\rho$ from Eqs.~(\ref{grg20})-(\ref{grg21}) gives
\begin{align}\label{grg222}
\Big[1-3(1+w)\kappa\lambda\Big]\ddot{\delta}a-\Bigg\{\Big[1+3w-6(1+w)\kappa\lambda\Big]\delta a+3(1+w)\kappa^{2}\lambda\delta\lambda\Bigg\}\frac{k}{a^2}=0.
\end{align}
Now, we insert the ansatz (\ref{grg2}) into Eq.~(\ref{grg222}) and use solution (\ref{grg17}) to more simplify this equation. From ansatz (\ref{grg2}) one obtains $\delta\lambda=n(\lambda/a) \delta a$ thus $\kappa^{2}\lambda\delta\lambda=(n/a)(\kappa\lambda)^{2}\delta a$, which finally leads to
\begin{align}\label{grg233}
&\ddot{\delta}a+6nk\beta\frac{w+1}{3w-1}\delta a=0,\\
&\beta=\left(\frac{\kappa\lambda}{a}\right)^{2}\frac{1}{a}.\nonumber
\end{align}
We observe that for those values of $w$ parameter for which $\beta$ is always positive (which in turn implies that $\mathcal{S}>0$ ), the scalar perturbation $\delta a$ does not diverge provided that
\begin{align}\label{grg24}
k=1,~\left\{
\begin{array}{l}
\zeta<0,~n<0,~~-1<w<-\frac{1}{3}\\
\zeta>0,~n>0,~~w<-1\bigvee w>\frac{1}{3}\\
\zeta>0,~n<0,~~|w|<\frac{1}{3}
\end{array}
\right.
~~~{\mbox and}~~~
k=-1,~
\left\{
\begin{array}{l}
\zeta<0,~n>0,~~-1<w<-\frac{1}{3}\\
\zeta>0,~n>0,~~|w|<\frac{1}{3}\\
\zeta>0,~n<0,~~w<-1\bigvee w>\frac{1}{3}
\end{array}
\right.
\end{align}
$\forall n, |n|\neq 1/2, 1/4, 1/6,...,$ and 
\begin{align}\label{grg25}
k=1,~\left\{
\begin{array}{l}
n<0,~~-1<w<\frac{1}{3}\\
n>0,~~w<-1~\bigvee~w>\frac{1}{3}
\end{array}
\right.
~~~{\mbox and}~~~
k=-1,~
\left\{
\begin{array}{l}
n<0,~~w<-1~\bigvee~w>\frac{1}{3}\\
n>0,~~-1<w<\frac{1}{3}.
\end{array}
\right.
\end{align}
$\forall n, |n|=1/2, 1/4, 1/6,\ldots,$ independent of the sign of $\mathcal{S}$. We therefore observe that except for cases with $\zeta<0$ the resulted intervals overlap with~(\ref{grg12}) and~(\ref{grg13}). For $\zeta<0$ and the equation of state parameter within the interval $-1<w<-1/3$, the scalar perturbations diverge and thus the ES universe does not remain stable.
%
\subsection{Vector and tensor perturbations}
Our calculations show that the vector and tensor perturbations for a barotropic perfect fluid in GRG obey the same perturbation equations as those of GR in a FLRW background. For the vector modes the following equation for the perturbations hold
\begin{align}\label{grg26}
\dot{\varpi}+(1-3c^{2}_{s})H\varpi=0,
\end{align}
where the co-moving dimensionless vorticity along with the square of sound speed are defined as $\varpi_{a}=a\varpi$ and $c^{2}_{s}=dp/d\rho$ respectively. Therefore, since in the ES universe we have $H=0$, the vector perturbations do not experience any type of evolution and thus they stay in a frozen state throughout the evolution of the ES universe.  

The tensor perturbations for a barotropic perfect fluid in a FLRW background also follow the same equation as the GR case, i.e., 
\begin{align}\label{grg27}
\ddot{\Sigma}_{k}+3H\dot{\Sigma}_{k}+\left[\frac{K^2}{a^2}+2\frac{k^2}{a^2}-\frac{8\pi G}{3}(1+3w)\rho\right]\Sigma_{k}=0,
\end{align}
where the definition of co-moving dimensionless transverse-traceless shear $\Sigma_{ab}=a\sigma_{ab}$ has been used and $K$ is the co-moving index in Fourier space. Using Eq.~(\ref{grg5}) along with solution~(\ref{grg17}) in an ES universe we obtain
\begin{align}\label{grg28}
\ddot{\Sigma}_{k}+\frac{1}{a_{ES}^2}\left\{K^{2}+2k\left[1-\frac{3w+1}{3\alpha(w+1)}\right]\right\}\Sigma=0,
\end{align}
where $\alpha=\kappa/8 \pi G$. Thus, assuming $\mathcal{S}>0$ and ignoring particular amounts of $n$ for which $\mathcal{S}$ can assume negative values, in order to have a stable ES state against tensor perturbations, we must have
\begin{align}\label{grg29}
&\alpha>\alpha_{min}\\
&\alpha_{min}=\frac{3 w+1}{3 (w+1) \left(\frac{K^2}{2 k}+1\right)}\nonumber.
\end{align}

We therefore conclude that to have oscillatory behavior for the tensor perturbations, $\kappa$ cannot take arbitrary values. In fact, for each co-moving index $K$ there is a minimum value for $\alpha$ parameter.
\section{Concluding Remarks}\label{conc}
Investigations have shown that GRG potentially deserves wide scopes of studies to extract physical content of the theory. In an independent work the present authors have shown that the GRG can describe an accelerated expansion phase for the universe, compatible with varying gravitational coupling theories which the most well-known of them is the Dirac proposal~\cite{moradpour2021}. We have also investigated GRG from a dynamical system point of view~\cite{shabani2022}. The GRG via the unknown Rastall parameter makes possible various studies in different frameworks. \\
In the framework of GRG, we investigated here, Einstein static type solutions using ansatz (\ref{grg2}) for the Rastall parameter. Since, in an ES universe all time derivatives vanish (i.e., $H, \dot{R}=0, \ldots$), thus the functionality of the Rastall parameter can include only the scale factor and, the most straightforward case is the power-law function. Interestingly, we found that this particular form of Rastall parameter leads to stable ES solutions. The stability condition consists of some constraints on the model parameters, $\zeta$ and $n$, and also the equation of state parameter, $w$. Our study shows that ES solution can exist in both spatially closed and open universes, at least from mathematical viewpoint. 
\par
A natural question is the capability of GRG to predict a successful exit of ES universe to an inflationary state. This issue has also been dealt with in the present work. We graphically illustrated that such an exist is possible in the GRG framework by allowing the equation of state parameter to vary with time. It is shown that in case where $w$ varies from $|w|<\frac{1}{3}$ to $w\approx-1/3$ the universe does experience a transition from an eternal ES state to an everlasting inflationary phase, for $\zeta>0$ and $n<0$. For the same values of these parameters, a different type of parametrization for $w$, makes a transition from an ES state to another ES state with smaller radius.
\par
The stability of obtained ES solutions was also analyzed under scalar, vector and tensor perturbations. Because of the particular form of the radius of ES universe, which has been obtaind in our calculations, all investigations about the stability problem are divided in two classes; $i)$ ES solutions with $\forall n, |n|\neq 1/2, 1/4, 1/6,...$: in this case, we found that to avoid unbounded growth of the scalar perturbations similar conditions to those of Subsection~\ref{sec3.1} hold for $\zeta>0$. That is, there is a nearly overlap in the results of stability studies between two methods (the dynamical system approach and perturbing the equations of motion) when $\zeta>0$. ii) those solutions with $\forall n, |n|= 1/2, 1/4, 1/6,...$: in these cases the two approaches give rise to compatible results for different signs of $n$.

Our calculations show that the vector perturbations remain frozen in the ES universe, that is, there is a neutral stability under the vector perturbations. Tensor perturbations regularly oscillate about a center point provided the ratio $\kappa/8\pi G$ exceed from a minimum value which depends on $w$, the comoving index and the sign of spatial curvature parameter. 

\end{document}